# Electrostatic modification of infrared response in gated structures based on $VO_2$

M. M. Qazilbash[1], Z. Q. Li[1], V. Podzorov[2], M. Brehm[3], F. Keilmann[3], B. G. Chae[4], H. T. Kim[4], D. N. Basov[1]

[1]*Physics Department, University of California-San Diego, La Jolla, California 92093, USA.*

[2]*Department of Physics and Astronomy, Rutgers University, Piscataway, New Jersey 08854, USA.*

[3]*Abt. Molekulare Strukturbiologie, Max-Planck-Institut für Biochemie & Center for NanoScience,82152 Martinsried, München, Germany.*

[4]*IT Convergence & Components Lab, Electronics and Telecommunications Research Institute (ETRI), Daejeon 305-350, Korea.*

**Abstract:** We investigate the changes in the infrared response due to charge carriers introduced by electrostatic doping of the correlated insulator vanadium dioxide ($VO_2$) integrated in the architecture of the field effect transistor. Accumulation of holes at the $VO_2$ interface with the gate dielectric leads to an increase in infrared absorption. This phenomenon is observed only in the insulator-to-metal transition regime of $VO_2$ with coexisting metallic and insulating regions. We postulate that doped holes lead to the growth of the metallic islands thereby promoting percolation, an effect that persists upon removal of the applied gate voltage.

Transition metal oxides (TMO) have been at the focus of fundamental physics research for decades.[1] A number of recent advances in the field are encouraging in the context of practical oxide electronics components with unique characteristics.[2] Many interesting and important TMO systems are insulating in their stoichiometric form but can be turned highly conductive via chemical doping.[1] Electrostatic modification of electronic and optical properties offers an alternative to chemical doping.[3] Several TMOs have been integrated in gated structures in which carrier density and conductivity can be modulated electrostatically enabling reversible tuning of properties of host TMO materials.[3,4,5] Thus TMO-based gated devices are emerging as an appealing platform for both exploring and exploiting diverse electronic effects in these systems. Here we describe a gated structure with vanadium dioxide ($VO_2$) as an active semiconductor with tunable conductivity. We show that the electric field across the gate insulator leads to increased electromagnetic absorption in the infrared part of the spectrum. The observed effects are connected with the electronic phase separation in the vicinity of the temperature-driven insulator-to-metal transition (IMT) in $VO_2$.

Infrared (IR) spectroscopy offers unique experimental access to electronic effects occurring in gated structures. The energy scales accessible through IR measurements are ideally suited to probe dynamical characteristics of the electron and/or hole gas formed at the interface between an active semiconductor and a gate insulator. Following the pioneering experiments on IR response of Si-based MOSFETs,[6] recent IR work on electrostatic charge injection has mainly focused on organic field effect transistors[7,8,9] and graphene-based devices.[10,11] Here we report on IR studies of gated structures based on $VO_2$ which is a prototypical correlated electron system. $VO_2$ exhibits the IMT at $T \approx 60^{o}C$ featuring interesting interplay of electron-electron correlations and charge density wave effects.[1,12,13,14,15] There have been several reports on observations of the IMT in $VO_2$ films integrated in field effect transistor (FET) devices.[13,16,17] One common aspect of prior work is that fairly large source (S) to drain (D) currents have been employed in all these experiments. We note that our experiments differ from these earlier studies. Indeed, no drain-source current need be applied for infrared monitoring of electrostatically-doped charges thus ruling out unwanted effects associated with current-induced heating.

In the inset of Fig.1 we show a schematic of our $VO_2$ gated devices. $VO_2$ films were grown on $(10\bar{1}0)$ $Al_2O_3$ (sapphire) substrates by the sol-gel method. Details of the preparation method and characterization procedures are given in Ref.18. We utilized a 1.5 μm layer of parylene as the gate insulator. Parylene was deposited on $VO_2$ surface at room temperature as described elsewhere.[19] The maximum strength of the electric field in devices with parylene insulator is comparable to that of the electric field strength in metal/$SiO_2$/Si field effect structures whereas the leakage current does not exceed 1 pA. Finally, we used a 20 nm layer of Indium Tin Oxide (ITO) to fabricate a transparent gate essential for an infrared probe of the electrostatically induced effects in $VO_2$. In Fig.1 we show a transmission spectrum $Tm(\omega)$ of the device. The opaque region 230-1200 $cm^{-1}$ is due to the sapphire substrate. Below 230 $cm^{-1}$ sapphire becomes transparent. It is clear from the plot that a thin layer of ITO does not significantly compromise transparency of our device with $Tm(\omega)$ approaching 40% near the low frequency cut-off.

Data in Fig. 2 and its inset display modifications of the infrared response under the applied bias. It is customary to represent data in a plot of two-dimensional absorption of our gated device $A(V_{GS}, \omega, T) = 1-[Tm(V_{GS}, \omega, T)/Tm(V_{GS} = 0V, \omega, T)]$. Here we focus on the far-IR part of the spectrum. Detailed studies of the temperature-driven IMT that we performed for similar samples show that modifications of the electronic response associated with the formation of the metallic state are most prominent in this particular spectral range.[15,20] We start with data collected at 342 K at which voltage-induced changes are most significant. We observe a systematic increase of absorption under the applied bias. This result points to enhanced metallicity of $VO_2$ at the interface with the gate insulator. Importantly, only positive polarity applied to $VO_2$ ($V_{GS} < 0$) corresponding to hole injection in the $VO_2$ layer leads to modifications of absorption. Electron injection yields negligible changes in our data barely exceeding the signal-to-noise of our experimental setup. Unexpectedly, hole-induced absorption persists after the bias has been reduced to zero value. Absorption values are restored to conditions before the application of the bias once the structure is cooled down below 320 K.

Within the frequency window accessible to us we do not find substantial ω-dependence of the absorption spectra. Therefore, we have chosen to represent trends seen in our data in the form of absorption integral $N(V_{GS}, T) = \int A(V_{GS}, \omega, T) d\omega$ over a frequency range between 45 $cm^{-1}$ and 230 $cm^{-1}$ (see Fig. 3). Generally, voltage-induced changes of $N(V_{GS})$ in a gated structure with transparent electrodes are associated with the formation of *two* accumulation layers on both sides of the gate insulator arising from the capacitive effect.[9] Previously, we have systematically investigated charge injection in ITO fabricated from the same target. Knowing the carrier density of the films $n_{ITO}=5\times10^{20}$ $cm^{-3}$, the effective mass $m_{ITO}=0.5m_e$ of carriers and their scattering rate $1/\tau_{ITO}=1300$ $cm^{-1}$ , one can evaluate field-induced changes of absorption following the analysis outlined in Ref.9. The diamonds in Fig.3a represent the upper limit for integrated absorption due to ITO layer $N_{ITO}(V_{GS})$ inferred from this analysis. This latter contribution accounts for less than 20% of the total effect and varies linearly with voltage. Thus electrostatic modification of $VO_2$ dominates in the data presented in Fig.3a. The voltage dependence of absorption enhancement appears to increase somewhat faster than the linear power of gate voltage for $V_{GS} < 0$ i.e. positive bias of $VO_2$, and remains essentially unchanged for $V_{GS} > 0$ (see Fig.3a). Voltage-induced modifications also reveal a pronounced temperature dependence of $N(T)$(see Fig 3b). The impact of the gate bias is negligible at 295 K. As one approaches the IMT with increasing temperature, gate-induced changes gain increasing prominence and are maximized in the immediate vicinity of the IMT at 342 K (see Fig 3b). At higher temperatures the film enters the metallic regime due to temperature-induced macroscopic percolation where the impact of the applied gate bias is rapidly diminished.

Data in Fig.3 uncover a rather exotic character of electrostatic modifications of the IR response of $VO_2$. A capacitive model of an FET device implies that IR absorption due to the accumulation layer should increase linearly with the applied gate voltage and should recover its $V_{GS}=0$ value once the gate is no longer biased. These expectations are in accord with earlier experiments.[7,9] In structures with significant leakage through the gate insulator the increase of absorption with $V_{GS}$ may be slower than linear.[7] Clearly, the behavior of $VO_2$-based structures is at variance with all these earlier observations.

An insight into the physics of the observed effects is provided by the near-field IR microscopy of $VO_2$ films. [15] These near-field imaging studies uncovered that the transition occurs via nucleation of the nano-scale metallic puddles in the insulating $VO_2$ host. As temperature increases these puddles grow in size, percolate and eventually lead to a nominally homogeneous metallic film. It is instructive to correlate the formation of the metallic puddles with the voltage-induced enhancement of absorption. It is evident from Fig.4 that the voltage induced changes are most significant in the immediate vicinity of the percolation of metallic puddles. This finding indicates that the applied bias may be promoting percolation in $VO_2$ rather than leading to the formation of a nominally uniform accumulation layer similar to common FET devices. It appears likely that the application of $V_{GS}$ results in a redistribution of the hole density in the bulk of the film because the holes are attracted toward the gate insulator-$VO_2$ interface, leading to a higher hole density near the interface. This promotes the growth of conducting metallic puddles and/or formation of new puddles in the near-surface region of $VO_2$.

The persistent nature of the gating-enhanced absorption means that electrostatically-doped charges promote the IMT in $VO_2$. We note that the resistance measurements of the classic temperature-driven IMT show hysteretic behavior typically seen in first-order phase transitions indicating the reluctance of the metallic phase to switch back to the insulating phase.[18] For the same reason, once the electric field has tipped the balance towards the metallic state in certain spatial regions of the $VO_2$ film, the subsequent removal of the applied field does not restore the modified configuration. Persistent enhancement of the conductivity in TMOs are of immense interest in the context of phase control of this class of materials.[21,22,23] In vast majority of experiments, these effects were triggered by photo-excitation in the visible or x-ray part of the spectrum. Persistent electro-absorption in $VO_2$ demonstrated in our work shares certain similarities with persistent photoconductivity in manganites. Indeed, both classes of materials demonstrate persistent effects in the regime of electronic phase separation where the insulating phase also reveals charge ordering.

In the IMT regime of $VO_2$, the insulating and metallic phases coexist and a small perturbation can lead to large changes because the perturbation favors one phase over the other. This is a hallmark of complex systems in which several competing interactions are at work and thereby lead to phase transitions and phase coexistence. [24] We emphasize that it is the accumulation of holes at the $VO_2$ interface with the gate dielectric, rather than electrons, that leads to the transformation of the insulating regions to metallic regions. This result is consistent with the general notion that charge transport in a Mott insulator is blocked by on-site Coulomb repulsion but becomes less energetically costly once holes are introduced into the system. Specific aspects of hole-induced IMT in $VO_2$ have been analyzed in Ref.[12,13]. Thus indirectly our experiments lend support to the dominant role of correlations amongst charge carriers in the IMT phenomenon of $VO_2$.

We thank Sun Jin Yun for a careful reading of the manuscript. We gratefully acknowledge discussions with Amos Sharoni and Ivan K. Schuller. This work was supported by the US Department of Energy and by ETRI.

**Figure Legends**

**Fig. 1**: Transmission of a $VO_2$-based device at $T = 295$ K. The cross-section of the device is shown in the inset. The thicknesses of the various layers in the device are as follows: $Al_2O_3$ substrate (0.4 mm), $VO_2$ (100 nm), thermally evaporated Ag contacts (20 nm), parylene (1.5 μm), and ITO (20 nm). The separation between the two Ag contacts is about 8 mm.

**Fig. 2**: The absorption $A(\omega)$ is plotted as a function of frequency at $T = 342$ K with increasing gate voltage $V_{GS}$ up to -300 V and then for decreasing $V_{GS}$ down to 0 V. Inset: The absorption $A(\omega)$ of $VO_2$ is plotted as a function of frequency at $T = 342$ K. The spectrum at $V_{GS} = 0$V, $t = 5$ min was obtained after a delay of five minutes after the spectrum $V_{GS} = 0$V, $t = 0$ min to show the stability of the spectrum at a fixed temperature.

**Fig.3**: (a) Integrated absorption $N(V_{GS})$ of $VO_2$ and ITO (squares) at $T$=342 K is plotted as a function of $V_{GS}$ for hole accumulation (negative axis) and electron accumulation (positive axis) in $VO_2$. The calculated integrated absorption $N(V_{GS})$ for ITO alone (diamonds) is plotted as a function of $V_{GS}$.(b) $N(T)$ of $VO_2$ and ITO is plotted for various temperatures for $V_{GS} = -300$V (hole accumulation in $VO_2$).

**Fig. 4**: The resistance of the $VO_2$ film is plotted with increasing temperature. Also shown are maps of near-field scattering amplitude of a $VO_2$ film for three temperatures (from left to right): insulating regime, phase coexistence, and rutile metal. The scanned area is 2 μm by 2 μm. The near-field scattering amplitude is higher (light blue and white) for the metallic regions compared to the insulating phase (dark blue).

Fig.1

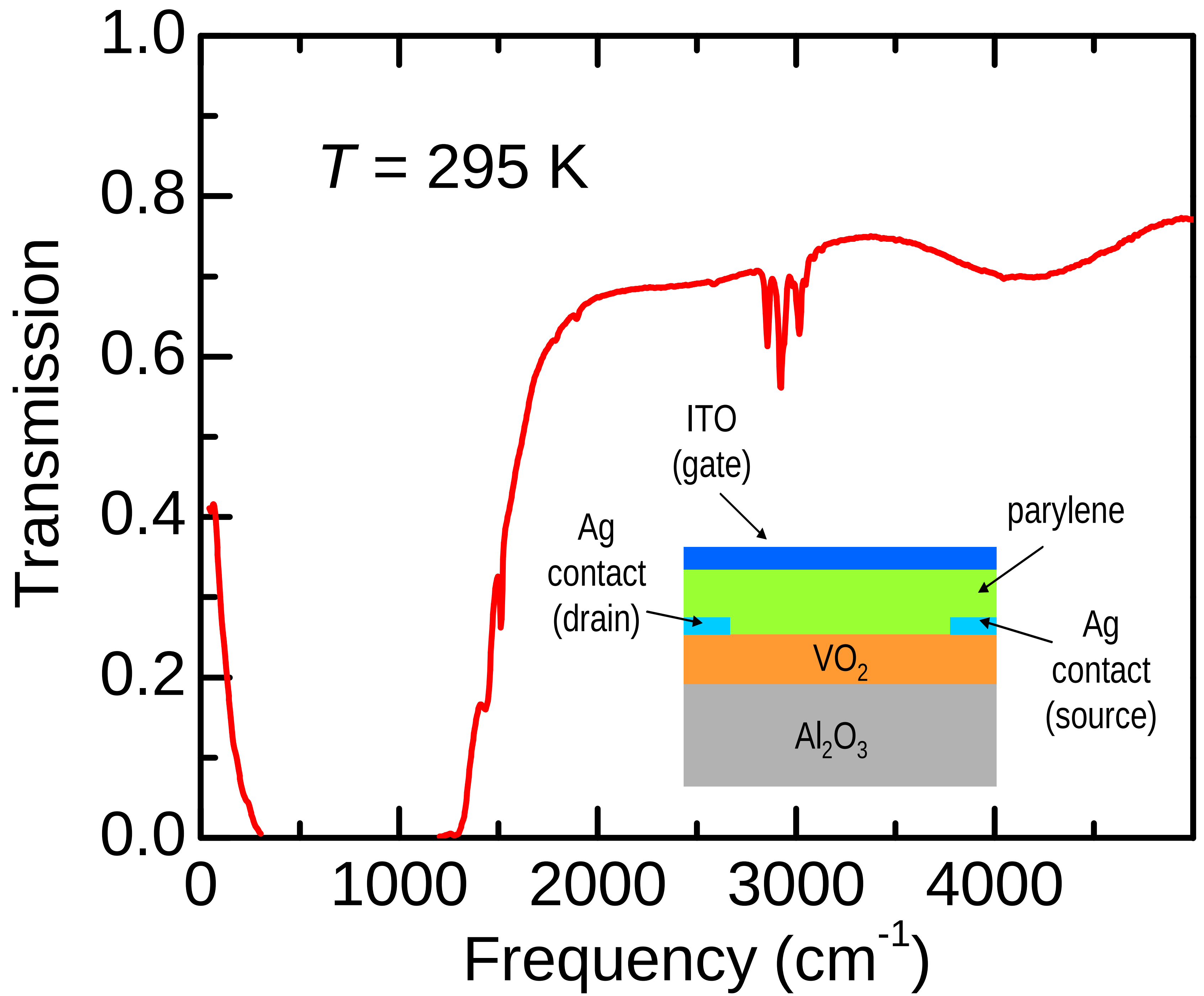

Fig. 2

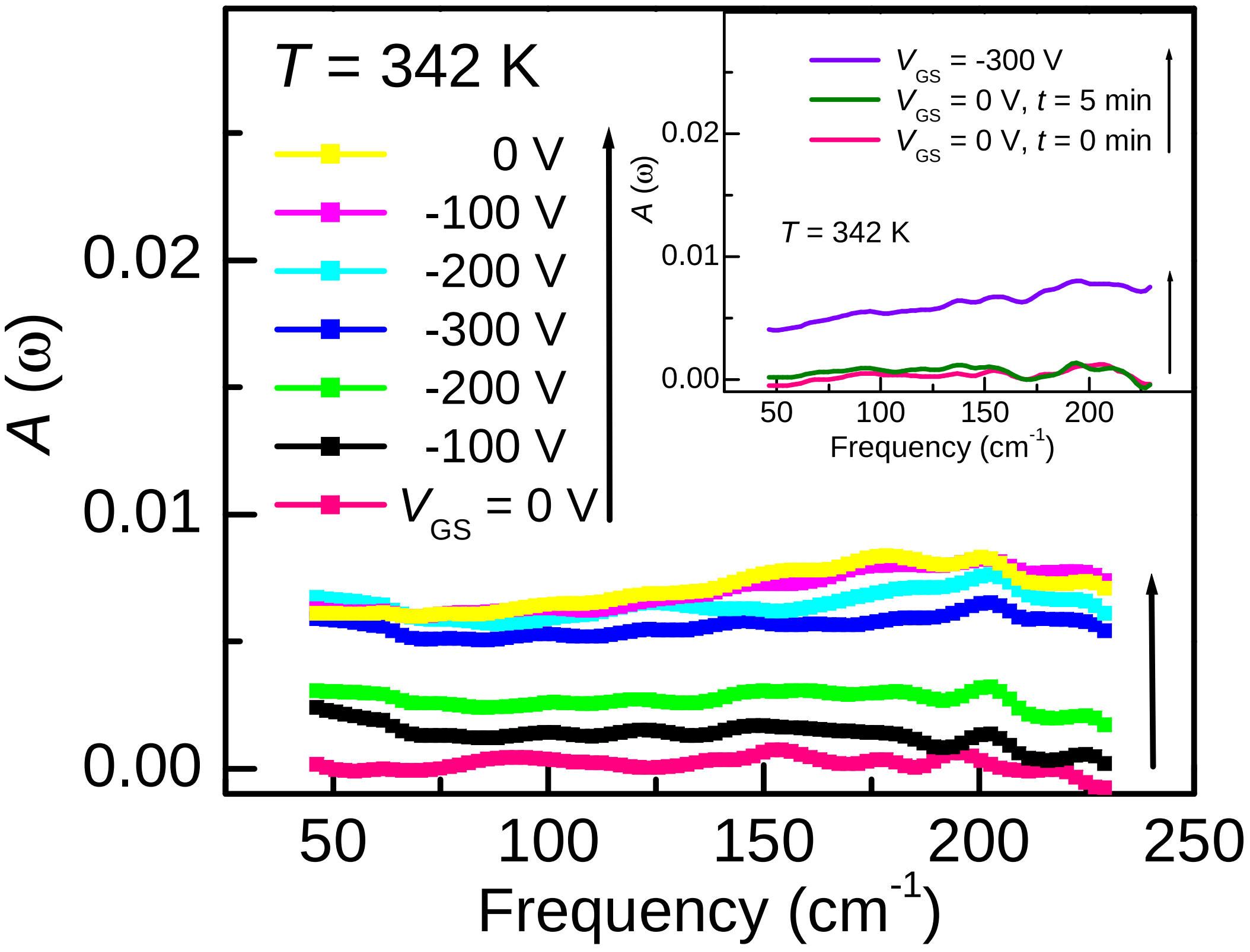

T = 342 K
0 V
-100 V
-200 V
-300 V
-200 V
-100 V
V_GS = 0 V
0.02
0.01
0.00
A (ω)
50
100
150
200
250
Frequency (cm^-1)
V_GS = -300 V
V_GS = 0 V, t = 5 min
V_GS = 0 V, t = 0 min
T = 342 K
0.02
0.01
0.00
A (ω)
50
100
150
200
Frequency (cm^-1)

Fig. 3

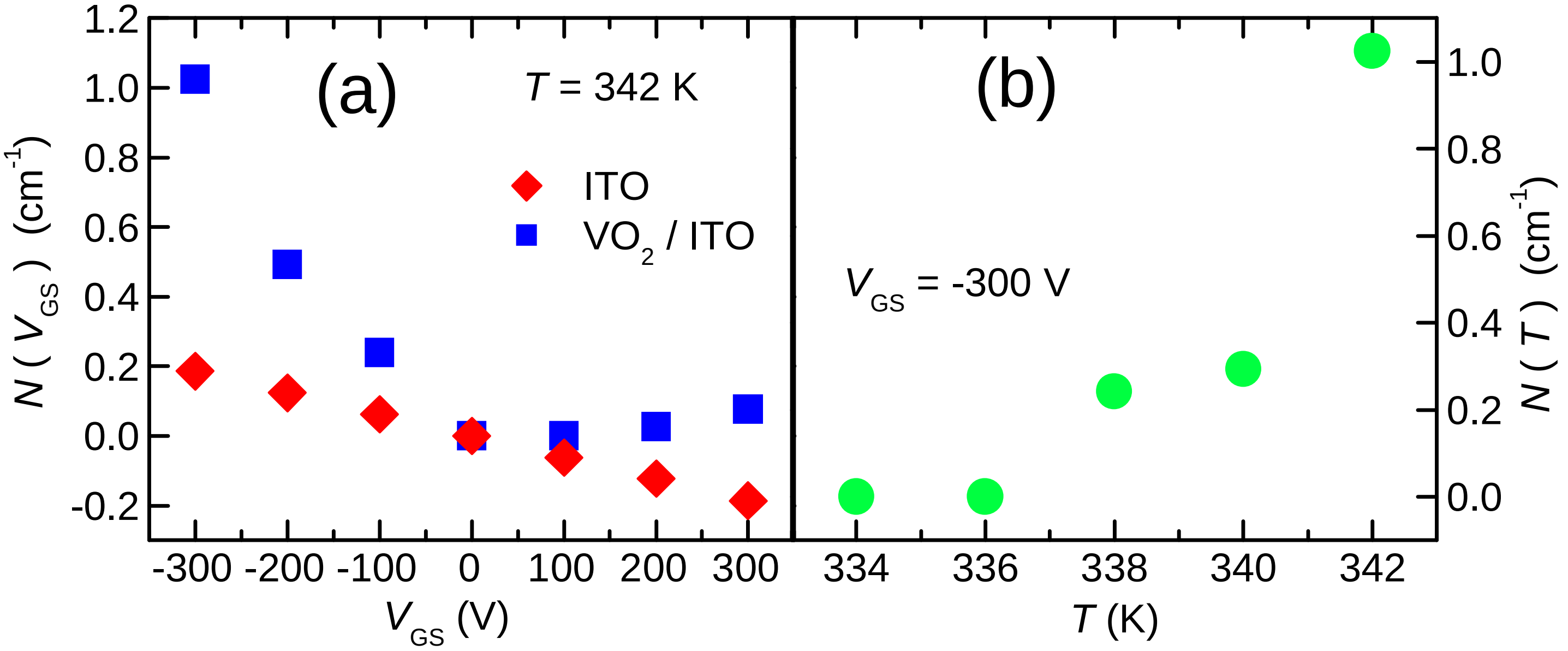

(a)
T = 342 K
ITO
VO2 / ITO
N ( VGS ) (cm-1)
1.2
1.0
0.8
0.6
0.4
0.2
0.0
-0.2
-300 -200 -100 0 100 200 300
VGS (V)
(b)
VGS = -300 V
N ( T ) (cm-1)
1.0
0.8
0.6
0.4
0.2
0.0
334 336 338 340 342
T (K)

Fig. 4

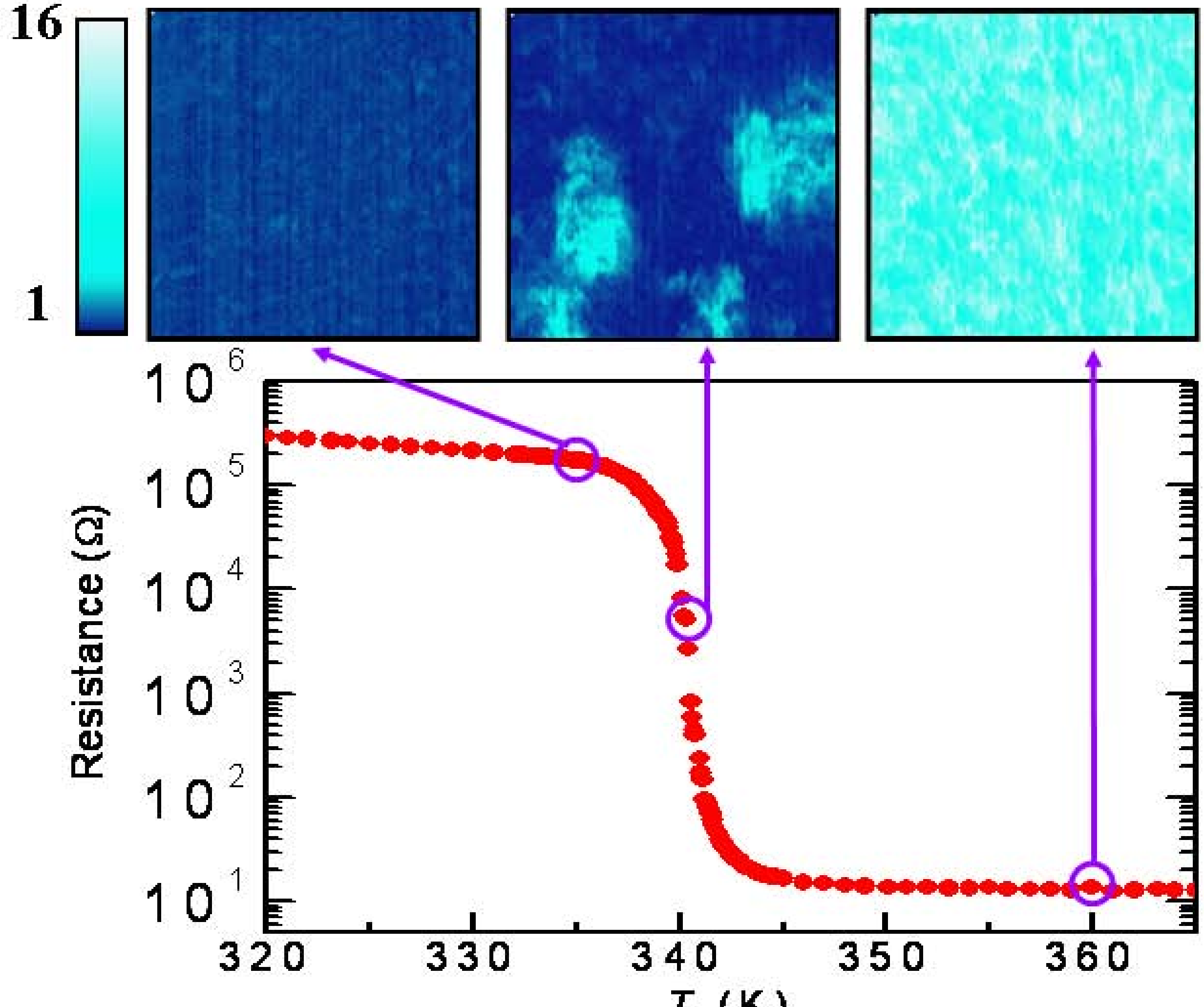